\documentclass[aps,prl,twocolumn,showpacs,groupedaddress]{revtex4-1}
\usepackage[latin9]{inputenc}
\usepackage{float}
\usepackage{amsmath}
\usepackage{graphicx}

\makeatletter
\usepackage{pifont}
\usepackage{amsfonts}
\usepackage{subfigure}
\usepackage{booktabs}
\usepackage{setspace}
\usepackage{threeparttable}
\usepackage{enumerate}
\usepackage{wasysym}
\usepackage{xspace}
\usepackage[T1]{fontenc}
\usepackage{textcomp}
\usepackage{txfonts}
\usepackage{bm}
\usepackage{graphicx,color,epsfig}
\usepackage{ulem}

\makeatother

\begin{document}

\title{Pairing from strong repulsion in triangular lattice Hubbard model}

\author{Shang-Shun Zhang$^{1}$, Wei Zhu$^{2}$ and Cristian D. Batista$^{1,3}$}

\address{$^1$Department of Physics and Astronomy, University of Tennessee, Knoxville,
Tennessee 37996-1200, USA}
\address{$^2$Theoretical Division, T-4 and CNLS, Los Alamos National Laboratory, Los Alamos, New Mexico 87545, USA}
\address{$^3$Quantum Condensed Matter Division and Shull-Wollan Center, Oak Ridge National Laboratory, Oak Ridge, Tennessee 37831, USA}

\begin{abstract}
We propose a paring mechanism between holes in the dilute limit of doped  frustrated Mott insulators.
Hole pairing  arises from a hole-hole-magnon three-body bound state.
This pairing mechanism has its roots on single-hole kinetic energy frustration, which favors
antiferromagnetic (AFM) correlations around the hole. We demonstrate that the AFM polaron (hole-magnon bound state) produced by a single-hole  propagating on a field-induced polarized background  is strong enough to bind a a second hole.
The effective interaction between these three-body bound states is repulsive, implying that this pairing mechanism is  relevant for superconductivity.
\end{abstract}
\maketitle

The long-sought  resonating valence bond  (RVB)  superconductor~\cite{Anderson87}, based on P. W.  Anderson's proposal  for describing the ground state of the antiferromagnetic (AFM)  triangular Heisenberg model~\cite{Anderson1973},
was the germ of two fundamental ideas in modern condensed matter physics. On the one hand, it set the basis for the search of quantum spin liquid states in Mott insulators. On the other hand, it suggested a clear connection between geometric frustration and superconductivity. The new century is witnessing an explosion of works based on the first idea~\cite{Balents2010}. However, while the superconductivity found in two-dimensional CoO$_2$ layers~\cite{Takada03} triggered some efforts related to the second idea~\cite{Ogata03,Baskaran03,Kumar03,Tanaka06,Motrunich04}, the relationship between  frustration and superconductivity  remained much less explored.

Nagaoka's theorem reveals a striking interplay between magnetism and electronic kinetic energy  in slightly doped Mott insulators \cite{nagaoka1966ferromagnetism,mielke1991ferromagnetism}.
The theorem states that a single hole propagating on a $D$-dimensional ($D>1$) bipartite  lattice with infinite on-site Hubbard repulsion $U$  minimizes its kinetic energy in a
ferromagnetic (FM) background. This well-known result inspired Schrieffer et. al. \cite{Schrieffer1988Spin-bag} to propose the ``spin-bag" mechanism for  pairing in doped Mott insulators.
A single hole propagating on a bipartite antiferromagnet generates a FM polaron or ``spin-bag" inside which the hole is self-consistently trapped. Two holes are attracted by sharing a common bag. While this  idea was never confirmed for the square lattice Hubbard model, it provides an interesting angle on
understanding how attraction (pairing) can potentially arise from strong bare repulsion.

\begin{figure}[!h]
\includegraphics[scale=0.85]{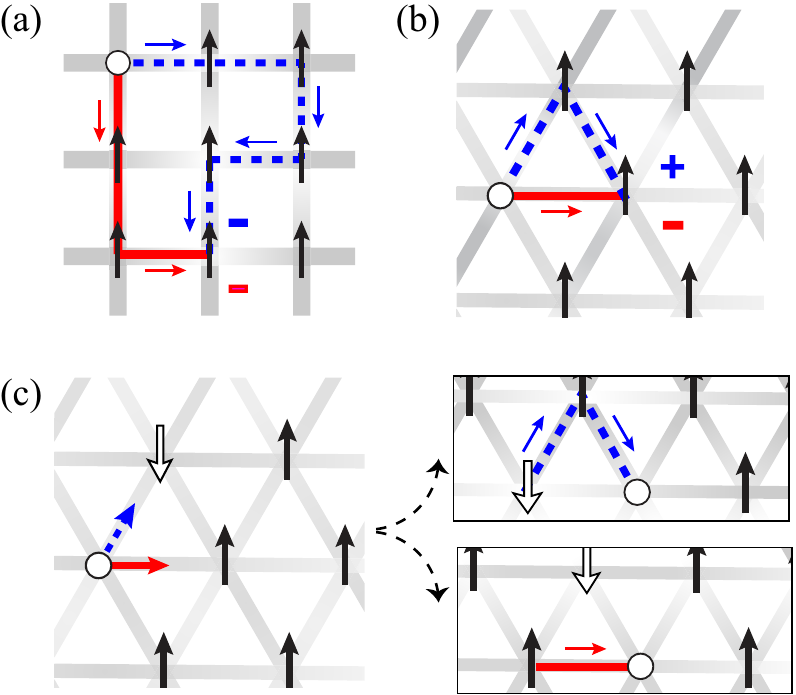}
\caption{(Color online) Single-hole propagation in fully polarized (a)  square  l and (b) triangular lattices with $t_1>0$. The signs indicate the optimal phase factor for each path assuming that the hopping matrix element is positive.
(c) Lack of destructive interference when one of the paths goes through a flipped spin.}
\label{fig:1hole}
\end{figure}

As shown in Fig.~\ref{fig:1hole}(a),  the kinetic energy of a single-hole on a bipartite lattice  is minimized ($- 4 \lvert t_1 \rvert$ for a square lattice with  nearest-neighbor (NN) hopping $t_1$) in a uniform (FM) background because  the interference between different paths connecting  two given points is always {\it constructive}. The situation can be very different for non-bipartite structures, such as the triangular lattice~\cite{haerter2005kinetic,Sposetti14,Lisandrini17}.
In this case, the single-hole kinetic energy is frustrated if the product of three hopping matrix elements  over the smallest closed loop of the lattice is {\it positive}.
For instance, the minimum kinetic energy of a single-hole on a triangular lattice is $-3 \lvert t_1 \rvert$ for a uniform FM background if the kinetic energy is frustrated ($t_1>0$) [see Fig.~\ref{fig:1hole}~(b)],
while it is $- 6|t_1|$ for the unfrustrated ($t_1<0$) case. Frustration arises from {\it destructive} interference between different paths. However, destructive interference can be avoided if the uniform FM background is replaced with a non-uniform state where one or more spins are flipped. As shown in Fig.~\ref{fig:1hole}(c), hole-paths  connecting two given points no longer interfere if one of the paths goes through a flipped spin~\cite{haerter2005kinetic,Sposetti14,Lisandrini17}.

In this Letter we demonstrate  that kinetic frustration is also the source of paring  between holes  near the fully polarized state induced by
an external or a molecular field $H$. Below the saturation field, $H_{\rm sat}$, it is energetically convenient to flip at least one spin. The single hole can then lower its kinetic energy by remaining close to a flipped spin. The resulting hole-magnon bound state, or 
 AFM {\it polaron}~\footnote{Note that the word {``polaron"} here is unrelated to lattice distortions, which are obviously absent in the purely electronic model that we are considering in this work.}, has a binding energy  $\sim - |t_1|/2$. In other words, the lowest single-polaron kinetic energy can reach a value as low as $-3.5 |t_1|$, which must be compared against the $-3|t_1|$ value obtained for a single hole (magnons have infinite mass for $U/|t_1| \to \infty$). Remarkably, the AFM  polaron mass, $m_{p} \simeq 10/|t_1|$, still has a moderate value. If a second hole is present, the strong hole-magnon attraction also leads to a {\it three-body bound state}, or AFM {\it   bipolaron}, which  still has an effective mass of order $10/|t_1|$. Moreover, our Density Matrix Renormalization Group (DMRG) results reveal a repulsive interaction between AFM bipolarons, implying that these composite Cooper pairs  should condense in the dilute limit.

We start by considering a Hubbard model  on a triangular lattice with NN hopping $t_1$ and the third NN hopping
$t_3$:
\begin{eqnarray}\label{eq:Hubbard}
{\cal H}_{H} &=& - t_{1}\sum_{\langle ij\rangle_{1} \sigma}c_{i \sigma}^{\dagger}c_{j \sigma}-t_{3}\sum_{\langle ij\rangle_{3} \sigma}c_{i \sigma}^{\dagger}c_{j \sigma} -\mu \sum_i n_i  \nonumber \\
&&+U \sum_i n_{i\uparrow}n_{i\downarrow} - H \sum_i S^z_i,
\end{eqnarray}
where $n_{\sigma}({\bm r}) =  c^{\dagger}_{{\bm r} \sigma} c_{{\bm r} \sigma}$,  $\mu$ the chemical potential  and $H$ the external magnetic field.

\begin{figure}[!t]
\includegraphics[scale=0.194]{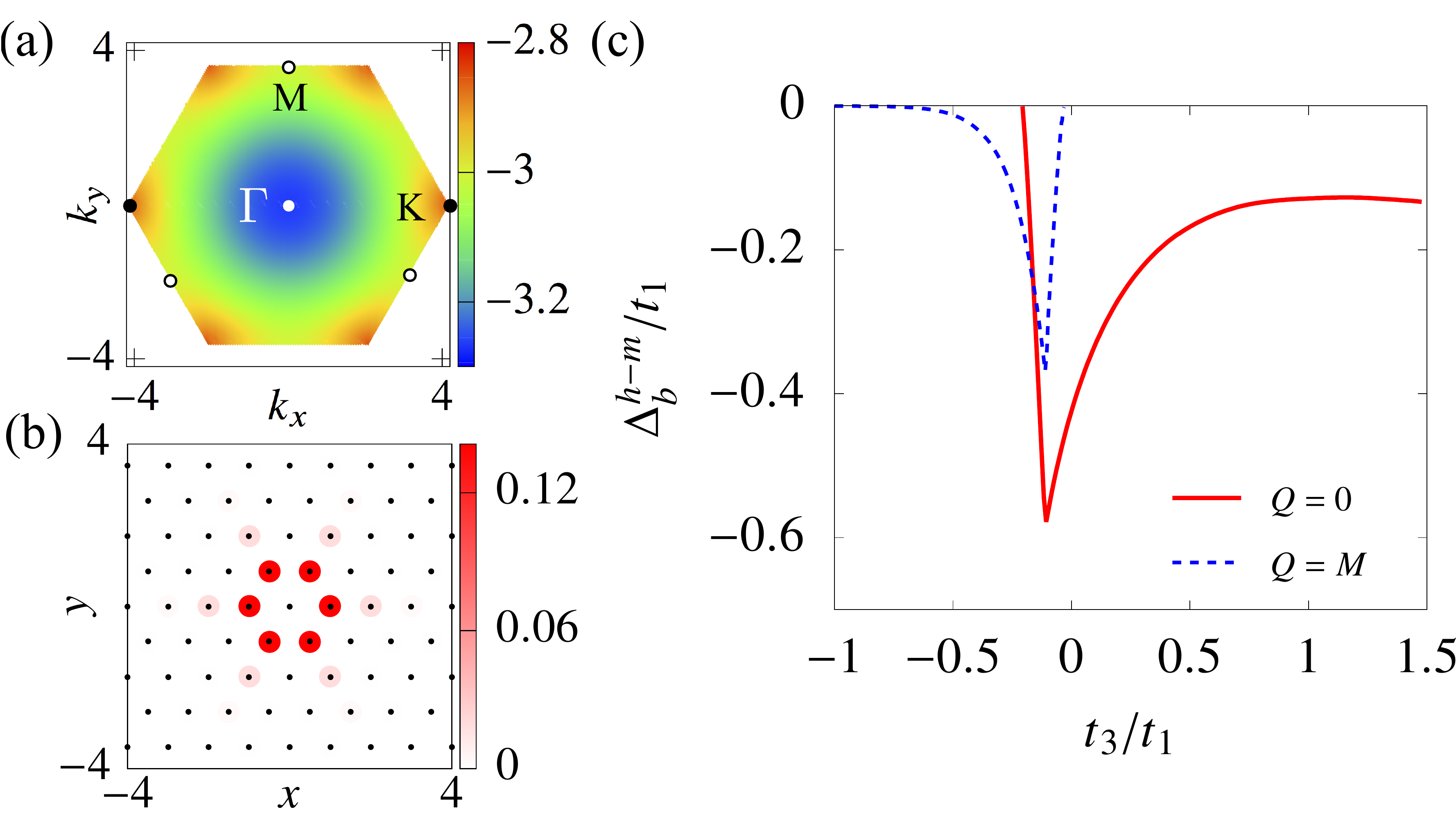}
\caption{(Color online) (a) Dispersion relation of the hole-magnon bound state with minimum at the $\Gamma$ point for $t_3=-0.1t_1$. 
(b) Correlation function between the hole and the magnon $ c^{h-m}({\bm r}) = \sum_{{\bm r}'} \langle n_{h} ({\bm r}-{\bm r}') n_{\downarrow}({\bm r}')\rangle$ for the lowest energy bound state at the $\Gamma$ point. 
(c) Binding energy of the hole-magnon bound state $\Delta^{h-m} = E_g (1H1S)- E_g (1H) - E_g (1S)$ as a function of $t_3/t_1$. 
}
\label{fig:1H1S}
\end{figure}


We will initially consider the $U / |t_1| \rightarrow \infty$ limit. 
The ground state is  fully polarized for $H>H_{\rm sat}$. 
In this regime the holes become non-interacting fermions with dispersion $\epsilon_{\bm k}=2t_1 [\cos(2 k_x/\sqrt{3}) + 2\cos(k_x/\sqrt{3}) \cos(k_y)]$ 
(note that the single-electron dispersion is $-\epsilon_{\bm k}$). 
The minimum energy of the single-hole spectrum  
$\epsilon_{\bm k}$ is $ \epsilon_{\bm 0}= -6\rvert t_1 \rvert $ for $t_1<0$ and  $ \epsilon_{\pm {\bm K}} = -3\rvert t_1 \rvert $ for $t_1>0$ (frustrated case) with ${\bm K}=({4\pi \over 3},0)$.

{\it AFM polaron.} The single-hole ground state is no longer fully polarized for $H< H_{\rm sat} = - \Delta^{hm}_{b}$, where $\Delta^{hm}_{b}$ is the hole-magnon binding state energy.
As anticipated, this bound state forms to suppress the destructive interference (frustration) of the single-hole motion. This idea can be illustrated with
a simple variational wave function for the relative coordinate ${\bm r}$ of the hole-magnon pair:
\begin{equation}
\psi(\bm{r})=\begin{cases}
\frac{\cos{\alpha}}{\sqrt{6}} e^{im\theta}, & \rvert\bm{r}\rvert=a,\\
\frac{\sin{\alpha}}{\sqrt{6}}  e^{i(m\theta+\phi)}, & \rvert\bm{r}\rvert=2a.
\end{cases}
\end{equation}
Here, $m=0,...,5$ is the crystal  angular momentum following from the $C_6$ symmetry of
${\cal H}_H$, $\theta =  n\pi / 3$ ($0\leq n \leq 5$) is the relative azimuthal angle and
$\phi$  is the phase difference  between the two particles separated by one and two lattice spaces $a$.
We are also assuming that the total momentum of the
two-particle system is equal to zero. 
Minimization of $E(\alpha)/t_{1}= (\cos{\alpha})^{2}\left(2\cos(\frac{m\pi}{3})+\cos(m\pi)\right)+ \sin{2 \alpha} \cos(\phi)$ for  
$t_1>0$ gives  $m=3, \phi = \pi$, $\cos{\alpha}=0.3907$  and $\tilde{E}_{\rm min} = -3.3028 t_1$, 
which is already quite close to the exact ground state energy $E_G = -3.4227 t_1$. 
The binding energy, $\text{min}_{\bf k \in \textit{BZ}}(\epsilon_{\bm k })-E_{G}$, is $0.4227 t_1$, 
indicating a strong effective attraction between the  magnon and the hole. 
The correlation function $c^{h-m}({\bm r}) = \sum_{{\bm r}'} \langle n_{h} ({\bm r}-{\bm r}') n_{\downarrow}({\bm r}')\rangle$,
shown in Fig.~\ref{fig:1H1S}~(b), reveals the spatial distribution of the magnon around the hole in the exact ground state.

The lowest energy magnon-hole pair  can also have finite center of mass momentum. 
Fig.~\ref{fig:1H1S}~(c) shows  the exact binding energy, $\Delta^{h-m}_b=E(1H1S)-E(1H)-E(1S)$, as a function of  $t_3/|t_1|$ that results from solving the
Lippmann-Schwinger (LS) equation in the thermodynamic limit~\cite{SM}. 
For  $t_3<-0.1615 t_1$, the lowest energy bound state is at the ${\bm M}$ point of the  Brillouin zone (BZ). 
The center of mass momentum of the ground state moves from the M to the $\Gamma$  point for  $t_3 > -0.1615 \rvert t_1 \rvert$.  
A positive $t_3$ does not change the nature of the bound state, which smoothly crosses over into another limit dominated by $t_3$ \footnote{For $t_1=0$ and $t_3>0$, 
the triangular lattice is divided into  four decoupled triangular sublattices with  lattice constant $2a$. After rescaling the length scale by a factor $1/\sqrt{3}$, 
each sublattice becomes the same as the original triangular lattice with  NN hopping. A small positive $t_1$ term couples the two sublattices and yields a ${\bm Q}=0$ bound state.}.

\begin{figure}[t!]
\includegraphics[scale=0.6]{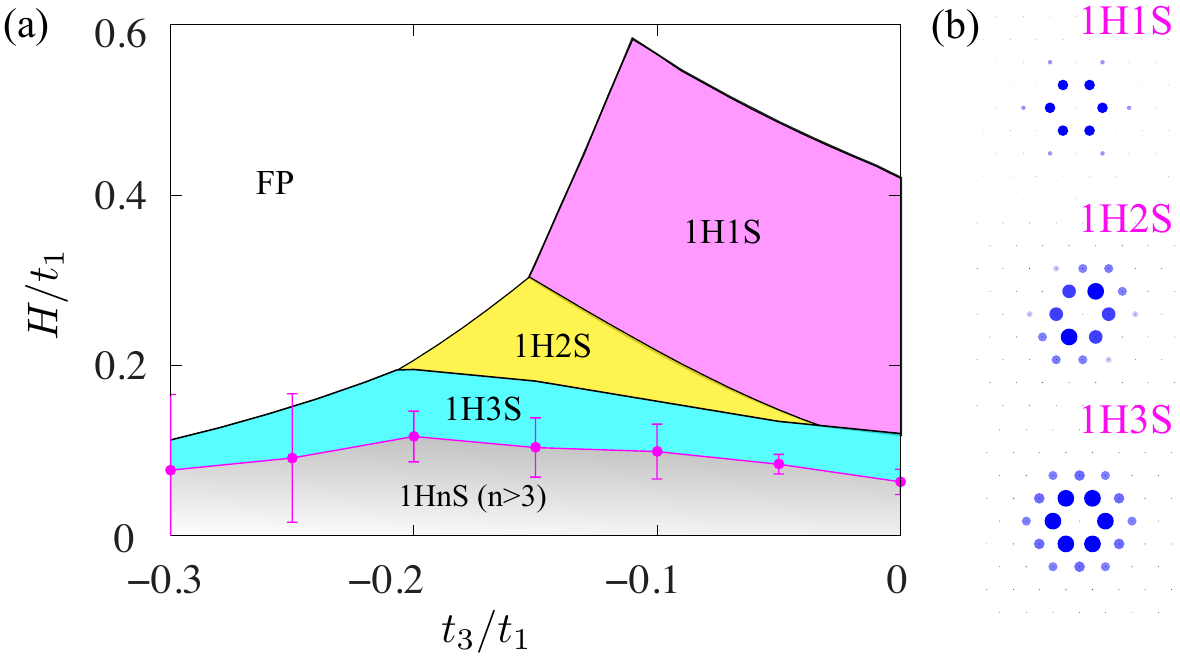}
\caption{(Color online)  (a) Phase diagram of one hole system on the $t_3-H$ plane with $t_1>0$. The grey region has stronger finite size effects. (b) Correlation function between the hole and the magnon $c^{h-m}({\bm r}) $  for different number of flipped spins ($t_3=-0.1t_1$).  }
\label{fig:pd1h}
\end{figure}



From now on, we will use the notation $mHnS$ to denote states with $m$ holes and $n$ flipped spins. 
We will consider  bound states of one hole ($m=1$) and $n \geq1$ flipped spins.  
Fig.~\ref{fig:pd1h}~(a) shows the phase diagram as a function of magnetic field and $t_3/t_1$. 
 The $1H1S$ is stable over a relatively large window of magnetic field values for small $|t_3|/t_1 \lesssim 0.1$. 
The number of magnons bounded to the hole increases continuously upon further decreasing the field.
The critical field for the transition into a $1HnS$ state decreases rapidly with $n$ because the  binding energy of the $n$th magnon goes asymptotically to zero for large $n$. 
This AFM polaron state is then expected to evolve smoothly into the long range AFM ordering found in Ref.~\cite{haerter2005kinetic} for $h \to 0$ ($n \to \infty$) because the radius of the AFM polaron (AFM correlation length) diverges. 
Fig.~\ref{fig:pd1h}~(b) shows the  evolution of the correlation function $ c^{h-m}({\bm r})$ as a function of $n$ for $n=1,2,3$.
For $n \leq 3$, the radius of the AFM polaron turns out to be significantly smaller than 
the linear size of the biggest lattices that enable exact diagonalization (ED) of ${\cal H}_{H}$.

\textit{Hole paring---} An important consequence of the  effective hole-magnon attraction is the possibility of indirect hole-hole pairing via formation of a {\it three-body bound state of two holes and one magnon} ($2H1S$). 
This  state can be regarded as an  AFM ``bipolaron" or ``spin-bag": the two holes
share the same AFM region to lower their kinetic energy at a minimum Zeeman energy cost.
Its wave function is also obtained by solving the LS equation in the thermodynamic limit~\cite{SM}, 
which provides a verification for the size effects of ED and DMRG calculations on finite lattices~\cite{SM}.  

\begin{figure}[!t]
\includegraphics[scale=0.35]{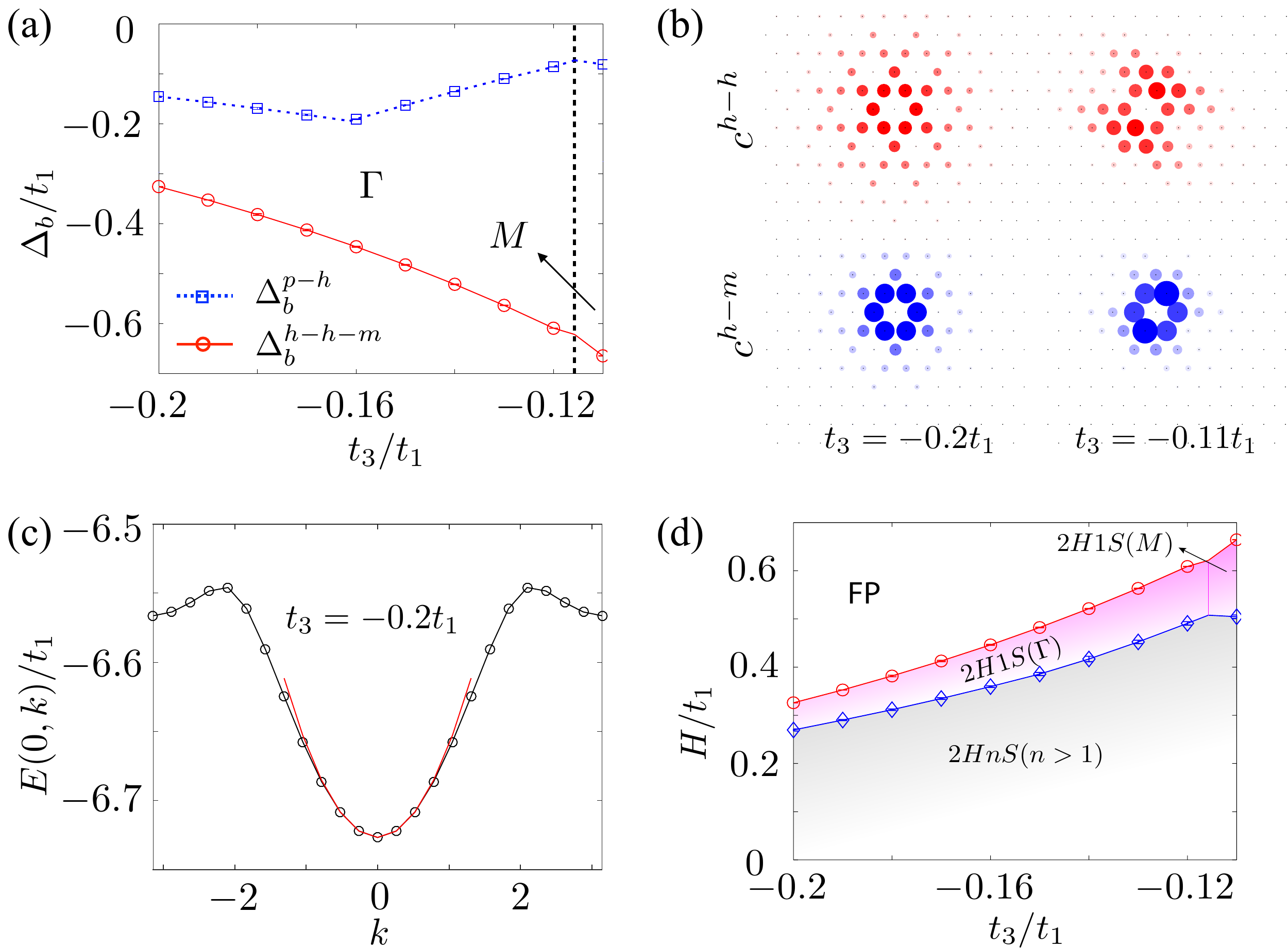}
\caption{(Color online)  (a) 
Binding energies $\Delta_b^{p-h}= E_g(2H1S) - E_g(1H1S)- E_g(1H)$ and 
$\Delta_b^{h-h-m}= E_g(2H1S) - 2 E_g(1H)- E_g(1S)$ as a function of $t_3/t_1$. The two kinks of $\Delta_b^{p-h}$ arise from a change in the center of mass momentum of the AFM polaron (at $t_3=0.175 t_1$) and the AFM bipolaron ( at $t_3=0.11 t_1$).
(b) Hole-hole, $ c^{h-h}({\bm r})$ and hole-magnon, $ c^{h-m}({\bm r})$, correlation functions  for different values of $t_3/t_1$ ($L=26$).
(c) Dispersion relation of the bipolaron ($2H1S$ bound state) for $t_3 = -0.2 t_1$ obtained from solving the LS equation (line) and from ED of an $L \times L$ lattice with $L=24$ (circles). The effective mass, $m_{\rm 2H1S}$, is extracted by fitting the the long wavelength region with a parabola. (d) Stability region of $2H1S$ state in the $ H-t_3/t_1$ phase diagram ($U\to \infty$ limit). }
\label{fig:E1hns}
\end{figure}

The two holes and the magnon indeed form a tight bound state for $t_1>0$. The lowest energy $2H1S$  bound state has center of mass momentum ${\bm Q}={\bm 0}$ for $t_3 < -0.116 t_1$ ($t_1>0$). The binding energy between a $1H1S$ polaron and  a second hole is defined as
$\Delta^{p-h}=E_g(2H1S)-E_g( 1H1S)- E_g(1H)$. It is also useful to introduce the binding energy of the three-body bound state relative to three non-interacting particles: $\Delta^{h-h-m}=E_g(2H1S)- 2 E_g( 1H)- E_g(1S)$. Both binding energies are shown in  Fig.~\ref{fig:E1hns}~(a). The negative value of  $\Delta^{p-h}$  demonstrates the AFM bipolaron formation, as confirmed by  the hole-hole, $ c^{h-h}({\bm r}) = \sum_{{\bm r}'} \langle n_{h} ({\bm r}-{\bm r}') n_{h}({\bm r}')\rangle$, and the hole-magnon, $ c^{h-m}({\bm r}) $,  correlation functions shown in Fig.~\ref{fig:E1hns}~(b).  Fig.~\ref{fig:E1hns}~(c) includes the AFM
bipolaron dispersion relation for $t_3=-0.2 t_1$, from which we extract an effective mass $m_{\rm 2H1S} \simeq 9.91 t_1^{-1}$.
The center of mass momentum of the lowest energy bound state moves to the ${\bm M}$ point of the BZ
for $-0.116 t_1<t_3<-0.1 t_1$. However, the bandwidth $W_{2H1S} \simeq 0.0737t_1$ is significantly narrower in this regime.
Correspondingly, the effective mass is large and anisotropic: $m^{\parallel}_{\rm 2H1S} \simeq 23.3t_1^{-1}$ and $m^{\perp}_{\rm 2H1S}66.67t_1^{-1}$ for the parallel and perpendicular directions relative to the ${\bm M}$ point.

As shown in Fig.~\ref{fig:E1hns}~(d), a second spin flips and binds to the $2H1S$ bound state upon further lowering $H$. The critical field  for flipping this spin is $H^c_{2S2H} = - \Delta_b^{b-m}$, where $\Delta_b^{b-m}=E(2H2S)-E(2H1S)-E(1S)$ is the binding energy between the AFM bipolaron and the second magnon. The critical field boundary shown in Fig.~\ref{fig:E1hns}~(d) is obtained from finite size scaling of the ground state energy~\cite{SM}.

\begin{figure}[!t]
\includegraphics[scale=0.4]{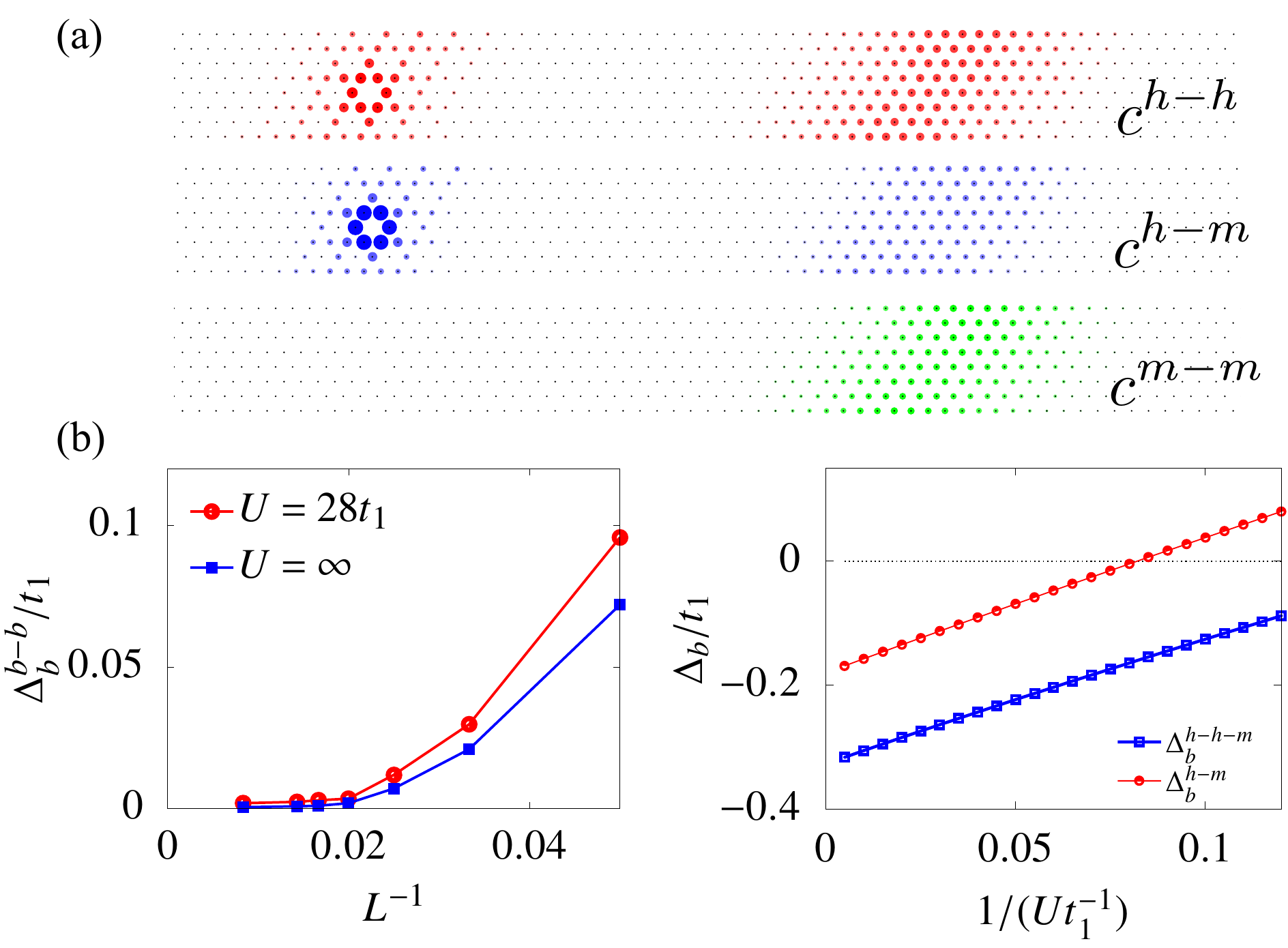}
\caption{(Color online) Correlation functions for four holes and two magnons: 
(a) Hole-hole, $ c^{h-h}({\bm r})$, hole-magnon, $ c^{h-m}({\bm r})$ and magnon-magnon, $ c^{m-m}({\bm r})$. 
The calculation is performed by DMRG simulation on $8\times 120$ lattice and setting  $t_1=1,t_3=-0.2,U=\infty$.
 This result indicates that the bipolarons ($2H1S$ bound states) are well separated. 
 (b) Finite size scaling of the binding energy between AFM bipolarons, $\Delta_b^{b-b}= E(4H2S)-2E(2H1S)$. 
 $\Delta_b^{b-b}$ is positive for finite $L$ and it extrapolates to zero in the $L\to \infty$ limit confirming the repulsive nature of the effective interaction. 
 (c) ED results: Binding energies $\Delta_b^{h-h-m}$ and $\Delta_b^{h-m}$  as a function of  $U$.}  
\label{fig:dmrg}
\end{figure}

{\it Interaction between AFM bipolarons---} Given that hole pairs are actually $2H1S$ bound states, 
we will further elucidate that  AFM bipolarons interact repulsively with each other, 
instead of forming larger bound states with multiple holes and magnons.
This is demonstrated  by solving the  six-body $4H2S$ problem.
Fig.~\ref{fig:dmrg}(a) shows the  hole-hole, hole-magnon and magnon-magnon correlation functions  for the ground state of the $4H2S$ system ($t_3/t_1 = -0.2$).
According to this result, the particles split into well separated $2H1S$  AFM bipolarons with the same correlation functions, $ c^{h-h}({\bm r})$ and $ c^{h-m}({\bm r})$, obtained for an individual bipolaron [see Fig.~\ref{fig:E1hns}(b)]. 
The magnon density-density correlation function, $ c^{m-m}({\bm r}) = \sum_{{\bm r}'} \langle n_{\downarrow} ({\bm r}-{\bm r}') n_{\downarrow}({\bm r}')\rangle$, confirms that each bipolaron contains one magnon.
The ground state energy of the $4H2S$ system $E_g(4H2S)$ equals twice the ground state energy of the $2H1S$ bound state, $E_g(4H2S)= 2 E_g(2H1S)$,  within an error of order $10^{-4}t_1$. In addition, as shown in Fig.~\ref{fig:dmrg}(b) for $t_3/t_1 = -0.2$, $\Delta_b^{b-b}= E_g(4H2S)- 2 E_g(2H1S)$ is positive for finite $L$ and it extrapolates to zero in the $L\to \infty$ limit, confirming the repulsive nature of the effective interaction. 
We note, however, that the two AFM bipolarons form a bound state when $t_3/t_1$ approaches  $-0.1$, i.e., in the region of strongest hole-magnon pairing  according to Figs.~\ref{fig:1H1S}(c) and~\ref{fig:E1hns}(a). However, as we discuss below, the interaction between AFM bipolarons becomes also repulsive in this region  for a finite $U  \lesssim 20 |t_1|$.


\textit{Effect of spin exchange---} 
Our next step is to analyze the effect of a finite, but still large, $U/|t_1| \gg 1$. The low-energy sector of the Hubbard model is now described by the $t-J$ model:
\begin{eqnarray}
{\cal H}_{t-J} &=& -t_{1}\sum_{\langle ij\rangle_{1}} {\tilde c}_{i \sigma}^{\dagger} {\tilde c}_{j \sigma}-t_{3}\sum_{\langle ij\rangle_{3}} {\tilde c}_{i \sigma}^{\dagger} {\tilde c}_{j \sigma} -\mu \sum	_i n_i  \nonumber \\
&&\!\!\!\!\!\!\!\!\!\!\!\!+J_{1}\sum_{\langle ij\rangle_{1}}\bm{S}_{i}\cdot\bm{S}_{j} +J_{3}\sum_{\langle ij\rangle_{3}}\bm{S}_{i}\cdot\bm{S}_{j} - H \sum_i S^z_i \nonumber \\
&&\!\!\!\!\!\!\!\!\!\!\!\! - \sum_{ijk \sigma} ^{i\neq k} { t_{ij} t_{jk} \over 2U} {\tilde c}_{i\sigma}^{\dagger} {\tilde c}_{k\sigma}n_{j\bar{\sigma}} + \sum_{ijk \sigma} ^{i\neq k} { t_{ij}t_{jk} \over 2U} {\tilde c}_{i\sigma}^{\dagger} {\tilde c}_{k\bar{\sigma}} {\tilde c}_{j\bar{\sigma}}^{\dagger} {\tilde c}_{j\sigma}.
\end{eqnarray}
$J_{\nu}=4t_{\nu}^2/U$ ($\nu=1,3$) and ${\tilde c}_{i \sigma}^{(\dagger)}$ are annihilation (creation) operators of constrained fermions: ${\tilde c}_{i \sigma}^{\dagger} = {c}_{i {\bar \sigma}}^{\dagger} (1- {c}_{i {\bar \sigma}}^{\dagger} c_{i {\bar \sigma}}^{\;}) $. The $XY$ part of the AFM exchange interactions, $J_1$ and $J_3$, generate a finite magnon mass,  while the Ising part induces a repulsive interaction  between any pair of particles (magnons or holes).
Consequently,  a finite $U/|t_1|$  should reduce the binding energy of the AFM polaron and bipolaron bound states. Indeed, as shown in Fig.~\ref{fig:dmrg}(c),  the AFM polaron bound state disappears below a critical value of $U/|t_1| \simeq 12.2 $, implying that the effective hole-hole attraction decreases upon {\it reducing} the bare Coulomb repulsion $U$. 
Moreover, a finite $U/|t_1|$ increases the repulsion between AFM bipolarons [see Fig.~\ref{fig:dmrg}(b)].


\textit{Order parameter---} Our results indicate the existence of a stable gas of AFM bipolarons  for low hole concentration $\rho_h \ll 1$ and $H \lesssim H_{\rm sat}$. In a pure 2D scenario, the AFM bipolarons must undergo a Berezinskii-Kosterlitz-Thouless transition~\cite{berezinskii1971destruction,berezinskii1972,kosterlitz1973ordering} into a superfluid state at a transition temperature $T_{BKT}$ of order 
$\rho_h/m_{\rm 2H1S}$.
The real space superconducting (SC) order parameter is $\Delta = \langle h_i^{\dagger} h_{j}^{\dagger} S_{k}^{-}\rangle$, where $i,j,k $ are  neighboring sites. Given the three-body nature of the bound state, the phase $\theta$ of the order parameter $\Delta$ includes a charge and a spin contribution (the superfluid current carries both charge and spin). The Hubbard Hamiltonian in a magnetic field ${\bm H}= H {\hat z}$ has a U(1)$\times$U(1) symmetry associated with the conservation of  the total charge and  $z$-component of the total spin. The phase $\theta$ is transformed into $\theta-\theta_s$ under a global spin rotation by an angle $\theta_s$ about the $z$-axis and into 
$\theta+2 \theta_c$ under a global charge rotation by an angle $\theta_c$ ($c^{\dagger}_{j \sigma} \to e^{-i \theta_c} c^{\dagger}_{j \sigma}$). In other words, the condensate is still invariant under the product of a spin rotation by $2\phi$ and a charge rotation by $\phi$ [U(1) subgroup]. This invariance implies lack of  long-range magnetic ordering in the condensate because the spin field can have arbitrary large phase fluctuations $\delta \theta_s$,  which are compensated by fluctuations of $\theta_c$.  Magnetic order can only take place via {\it single} magnon condensation. 

Finally, the pairing symmetry is determined by the irreducible representation of the single AFM bipolaron ($2H1S$) ground state.
For $t_3/t_1 = -0.2$ ($t_1>0$), the wave function of the $2H1S$ bound state has  zero total momentum and it belongs to  the $B_2$ representation of the $D_6$ space group  ($f_2$-wave)~\cite{yanase2005multi,nisikawa2004possible}.

\textit{Discussion---}
The possibility of generating an effective  attraction between electrons out of the bare Coulomb repulsion is a long sought-after goal of the condensed matter community~\cite{Little64,Ginzburg64,Kohn65,Fay68,Ginzburg76,Hirsch85,Micnas90,Chubukov92,Chubukov93,Raikh96,Hlubina99,Miraz04,Isaev10,Raghu10,Alexandrov11,Raghu11,Raghu12,Hamo16}. Here we have shown that magnons provide a strong glue in the infinitely repulsive limit of a slightly doped frustrated Mott insulator. The strongly attractive hole-magnon interaction is a manifestation of  the ``counter-Nagaoka" mechanism reported in Refs.~\cite{haerter2005kinetic,Sposetti14,Lisandrini17}: a single-hole can lower its kinetic energy by creating hole-magnon bound state (AFM polaron ).  The second hole binds to the polaron to lower its kinetic energy at a minimum Zeeman energy cost (AFM bipolaron). We have also verified that AFM bipolarons interact repulsively with each other. We note that these composite pairs have a  pure electronic origin and they are qualitatively different from the lattice bipolarons arising from a strong electron-phonon coupling~\cite{wellein1996polarons,verbist1990stability,emin1989formation}.

It is important to clarify that a saturation field of order $|t_1|/2$ is much higher than the maximum fields that can be generated in the laboratory. Moreover, such a large external field would produce a huge orbital effect that is  not included in our analysis. For charged systems, like electrons in a solid (Na$_x$CoO$_2$ is a well-known realization of a triangular lattice Hubbard model~\cite{Levi03}), this problem can be avoided by replacing the external field with a molecular field generated by  interaction  between the moments ${\bm S}_j$   and an insulating ferromagnetic layer.  For neutral systems, such as ultracold  two-component fermionic  gases of atoms~\cite{becker2010ultracold,struck2011quantum,struck2013engineering,bloch2008many,jaksch2005cold}, the orbital effect is not present and the system can be easily driven into the fully polarized state. Nevertheless, the main purpose of our analysis  is to understand how magnetic excitations can provide the glue for hole-hole pairing in the vicinity of a magnetic field induced AFM quantum critical point of the finite-$U$ Mott insulating state.
 Remarkably, we find that antiferromagnetism (single-magnon condensation) is suppressed by the  AFM bipolaron condensate (SC state) because  magnons do not condense individually, but as a component of a three-body bound state.
This simple mechanism then illustrates the competition between antiferromagnetism and superconductivity: magnons can either condense individually to form an AFM state or become part of an AFM bipolaron that condenses into a SC state.

\begin{acknowledgments}
{\it Acknowledgments.} We thank Zhentao Wang, Sriram Shastry and Andrey Chubukov 
for helpful discussions. C. D. B and S-S. Z. are supported by funding from the Lincoln Chair of Excellence in Physics.
W. Z. was supported by DOE National Nuclear Security Administration through Los Alamos National Laboratory LDRD Program.
\end{acknowledgments}

\clearpage

We  include analytical solutions of the  two and three-body bound state problems
via the Lippmann-Schwinger equation. For states  with more than three particles $N_p = N_h + N_s$ ($N_h$ is the number of holes and $N_s$ is the number of flipped spins relative to the fully polarized state), we used exact diagonalization (ED) and density matrix renormalization group (DMRG). We include a finite size scaling analysis of these results, as well as real space correlations functions revealing the repulsive nature of the interaction between antiferromagnetic (AFM) bipolarons.

\section{Analytic approach to the few body problem}

Below we derive the Lippmann-Schwinger equation in the $U \to \infty$ limit. The extension to 
 finite $U$ is straightforward.

\subsection{Hole-magnon bound state}

The wave function of the hole-magnon bound state, $\rvert\psi\rangle=\sum_{\bm{r}}\psi_{Q}(\bm{r})\frac{1}{\sqrt{N}}\sum_{\bm{R}}e^{i\bm{Q}\cdot\bm{R}}\rvert\bm{R},\bm{R}+\bm{r}\rangle$, can be obtained by solving the Lippmann-Schwinger equation:
\begin{eqnarray}
\psi_{Q}(\bm{r}) & = & \sum_{\nu}t_{\bm{e}_\nu}G_{\bm{Q},E}(\bm{r}+\bm{e}_{\nu})\psi_{Q}(\bm{e}_{\nu})e^{-i\bm{Q}\cdot\bm{e}_{\nu}}+VG_{\bm{Q},E}(\bm{r})\psi_{Q}(\bm{0}), \nonumber\\
&&  \label{eq:LSE}
\end{eqnarray}
where $\bm{e}_\nu$ are (bond) vectors connecting  nearest  and third-nearest-neighbor sites. The hard core constraint of spins and holes (a flipped spin and a hole cannot occupy the same site) is imposed by including an infinitely repulsive on-site interaction $V\rightarrow\infty$.
The hole-magnon Green's function is:
\begin{align}
G_{\bm{Q},E}(\bm{r}_{1}-\bm{r}_{2}) & =\langle\bm{Q},\bm{r}_{1}\rvert\frac{1}{E-{\cal H}_{0}+i\epsilon}\rvert\bm{Q},\bm{r}_{2}\rangle\\
 & =\int\frac{d^{2}\bm{k}}{(2\pi)^{2}}\frac{e^{i\bm{k}\cdot(\bm{r}_{1}-\bm{r}_{2})}}{E-(\xi_{\bm{Q}-\bm{k}}+\omega_{\bm{k}})+i\epsilon}.\label{eq:GFhm}
\end{align}
The   hole-magnon hard-core interaction implies
\begin{equation}
\sum_{\nu}t_{\bm{e}_\nu}G_{\bm{Q},E}(-\bm{e}_{nu})\psi_{Q}(\bm{e}_{\nu})e^{-i\bm{Q}\cdot\bm{e}_{\nu}}+VG_{\bm{Q},E}(\bm{0})\psi_{Q}(\bm{0})=0.
\end{equation}
After applying this condition,
the Lippmann-Schwinger equation~\eqref{eq:LSE} becomes
\begin{eqnarray}
\psi_{Q}(\bm{r}) \!= \!\!\! \sum_{\nu} \! t_{\bm{e}_\nu} \!  \left( G_{\bm{Q},E}(\bm{r}+\bm{e}_{\nu})\!-\! \frac{G_{\bm{Q},E}(\bm{r})}{G_{\bm{Q},E}(\bm{0})}
G_{\bm{Q},E}(-\bm{e}_{\nu})\ \right) \! \psi_{Q}(\bm{e}_{\nu})e^{-i\bm{Q}\cdot\bm{e}_{\nu}}.
\nonumber \\
\label{eq:LSE-1}
\end{eqnarray}
By setting $\bm{r}=\bm{e}_{\eta}$, we obtain twelve coupled linear
equations for $\psi_{Q}(\bm{e}_{\nu})$, which  determine  $\psi_{Q}(\bm{r})$ through Eq. (\ref{eq:LSE-1}). The
coefficients of the linear system of equations are computed by  using a numerical
integration method to evaluate the hole-magnon Green's
function given in Eq.~(\ref{eq:GFhm}).

\subsection{Antiferromagnetic Bipolaron}

The  wave function for two holes and one magnon is:
\begin{equation}
\rvert\psi\rangle=\sum_{\bm{r}_{1},\bm{r}_{2}}\psi_{Q}(\bm{r}_{1},\bm{r}_{2})\frac{1}{\sqrt{N}}\sum_{R}e^{i\bm{Q}\cdot\bm{R}}\rvert\bm{R,R+r}_{1},\bm{R}+\bm{r}_{2}\rangle.
\end{equation}
The fermionic statistics of holes implies $\psi_{Q}(\bm{r}_{1},\bm{r}_{2})=-\psi_{Q}(\bm{r}_{2},\bm{r}_{1})$. The two-hole wave function with total momentum of $\bm{Q}$
can be re-expressed as a function of the position $\bm{r}$ of one hole and the momentum $\bm{p}$ of the second hole:
\begin{align}
\psi_{\bm{Q}}(\bm{p},\bm{r}) & =\frac{1}{\sqrt{N}}\sum_{\bm{r^{\prime}}}e^{-i\bm{p}\cdot\bm{r^{\prime}}}\psi_{\bm {Q}}(\bm{r}^{\prime},\bm{r}).
\end{align}
The Lippmann-Schwinger equation becomes:
\begin{align}
\psi_{Q}(\bm{p},\bm{r}) & =2\sum_{\nu}t_{\nu}\int\frac{d^{2}\bm{k}}{(2\pi)^{2}}G_{\bm{Q}E}^{(3)}(\bm{p},\bm{r};\bm{k},-\bm{e}_{\nu})e^{-i(\bm{Q}-\bm{k})\cdot\bm{e}_{\nu}}\psi_{Q}(\bm{k},\bm{e}_{\nu})  \nonumber  \\
 & +2V\int\frac{d^{2}\bm{k}}{(2\pi)^{2}}G_{\bm{Q}E}^{(3)}(\bm{p},\bm{r};\bm{k},\bm{0})\psi_{Q}(\bm{k},\bm{0}),
\end{align}
where $G_{\bm{Q}E}^{(3)}$ is the non-interacting three-body Green's function of the non-interacting  hole-hole-magnon system:
\begin{align}
&G_{\bm{Q}E}^{(3)}(\bm{p}_{1},\bm{r}_{1};\bm{p}_{2},\bm{r}_{2}) 
= \langle\bm{Q};\bm{p}_{1},\bm{r}_{1}\rvert\frac{1}{E-{\cal H}_{0}+i\epsilon}\rvert\bm{Q};\bm{p}_{2},\bm{r}_{2}\rangle   \nonumber  \\
 =& 2\pi^{2}\delta_{\bm{p}_{1}\bm{p}_{2}}\int\frac{d^{2}\bm{k}}{(2\pi)^{2}}\frac{e^{-i\bm{k}\cdot(\bm{r}_{2}-\bm{r}_{1})}}{E-(\xi_{\bm{p}_{2}}+\xi_{\bm{k}}+\omega_{\bm{Q}-\bm{p}_{2}-\bm{k}})}   \nonumber  \\
 & -\frac{1}{2}\frac{e^{-i\left(\bm{p}_{1}\cdot\bm{r}_{2}-\bm{p}_{2}\cdot\bm{r}_{1}\right)}}{E-(\xi_{\bm{p}_{2}}+\xi_{\bm{p}_{1}}+\omega_{\bm{Q}-\bm{p}_{2}-\bm{p}_{1}})}.
\end{align}
The hard-core constraint gives a boundary condition at ${ \bm r}={\bm 0}$:
\begin{align}
0 & =2\sum_{\nu}t_{\nu}\int\frac{d^{2}\bm{k}}{(2\pi)^{2}}G_{\bm{Q}E}^{(3)}(\bm{p},\bm{0};\bm{k},-\bm{e}_{\nu})e^{-i(\bm{Q}-\bm{k})\cdot\bm{e}_{\nu}}\psi_{Q}(\bm{k},\bm{e}_{\nu})  \nonumber  \\
 & +2V\int\frac{d^{2}\bm{k}}{(2\pi)^{2}}G_{\bm{Q}E}^{(3)}(\bm{p},\bm{0};\bm{k},\bm{0})\psi_{Q}(\bm{k},\bm{0}),
\end{align}
By setting $\bm{r}=\bm{e}_{\eta}$, we obtain a system of twelve coupled integral
equations for the  functions $\psi_{Q}(\bm{p},\bm{e}_{\nu})$,
which in turn determine the three-body wave function $\psi_{Q}(\bm{p},\bm{r})$.

The other three-body bound state of one hole and two flipped spins (1H2S) is obtained in a similar way.

\section{Finite size effects in Exact Diagonalization simulation}

The binding energy and various correlation functions for $n$ holes and $m$ flipped spins ($nHmS$) are calculated by the Lanczos method on  finite  triangular lattices of $L\times L$ sites.
The finite size correction for bound states of linear size $\xi\ll L$ is exponentially small in $\xi/L$. 
This can be verified for the two-body and three-body bound states, whose solutions are obtained by solving the Lippmann-Schwinger equation  in the thermodynamic limit .
Finite size corrections become more important for  bound states composed of more than three particles because $\xi$ increases, while the maximum accessible lattice size  decreases.

\begin{figure}[!t]
\includegraphics[scale=0.25]{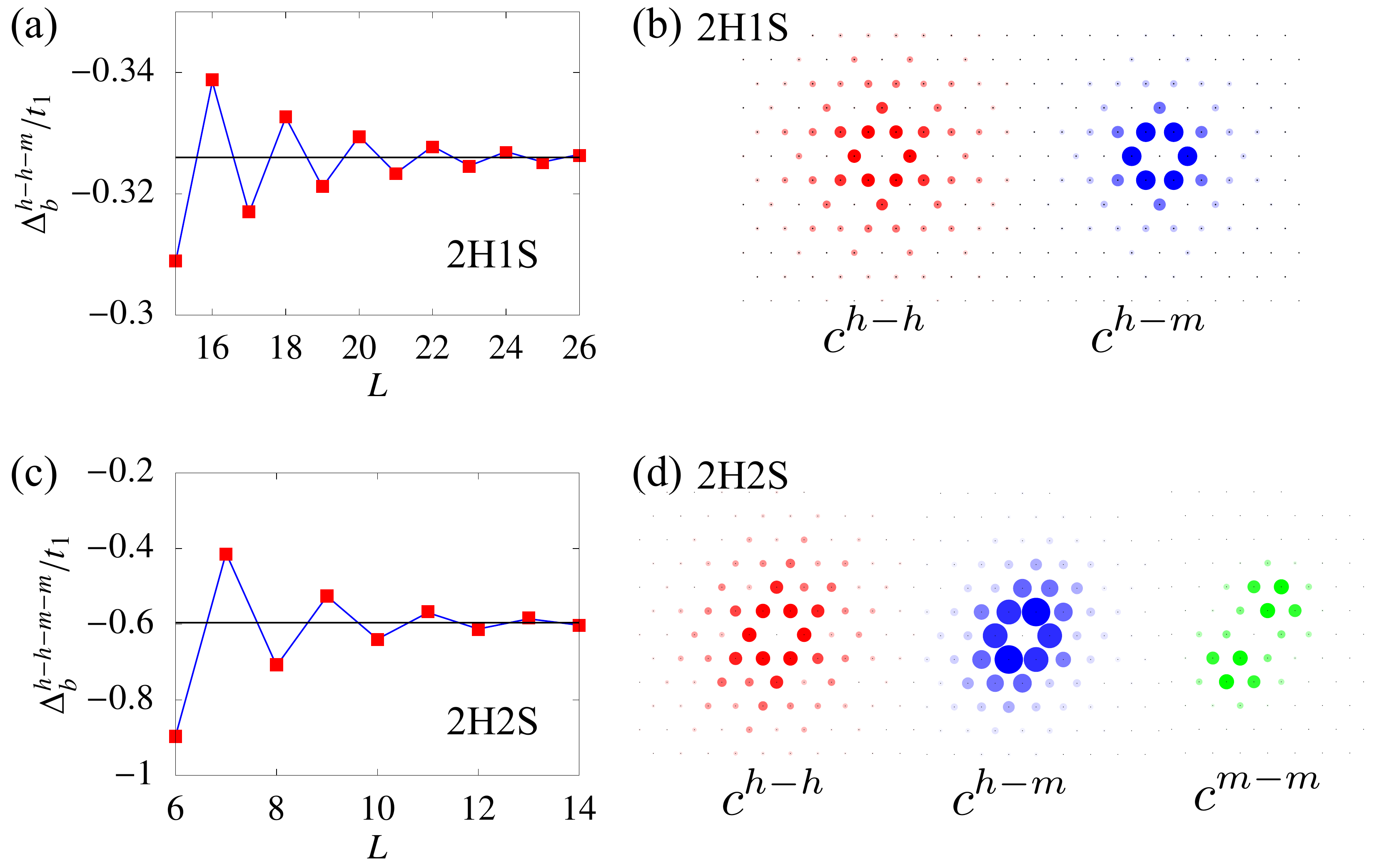}
\caption{(Color online) (a) Fit of size dependence of the ground state energy for $2H1S$ using Eq.~(\ref{fit}). (b) Hole-hole and hole-magnon density-density correlation functions obtained with ED for a lattice of  $26\times26$ sites. (c) Fit of size dependence of the ground state energy for $2H2S$ using Eq.~(\ref{fit}). (d) Hole-hole, hole-magnon and magnon-magnon density-density correlation functions obtained  for $2H2S$ on a lattice of  $26\times26$ sites. In all cases, the Hamiltonian parameters are: $t_1=1,t_3=-0.2,U=\infty$.}
\label{fig:fit}
\end{figure}

To extract the error due to finite size effects, we perform a finite size scaling analysis of the numerical data.
We fit the ground state energy $E_G(L)$ for a finite $L\times L$ lattice size using the  formula
\begin{equation}
\label{fit}
E_G(L)=E_G(\infty) + (-1)^L A \exp(-L/\xi), L\gg \xi.
\end{equation}
The factor $(-1)^L$ accounts an oscillation between even and odd linear system sizes, arising from kinetic energy frustration on finite lattices. As an example,
Fig.~\ref{fig:fit} shows  the fit of the ground state energy using Eq.~(\ref{fit}) for systems composed of $2H1S$ and 
$2H2S$, respectively. The data obeys Eq.~(\ref{fit}) very well, indicating that the system size is larger than $\xi$.
Within a  confidence interval of $95\%$, the binding energy for the $2H1S$ bound state is $E_G(\infty)=-0.32591 \pm 1.1647\times 10^{-4}$, while the binding energy for the $2H2S$ bound  state is $-0.59538 \pm 0.00205$. This is the method that we used to extract the binding energies reported in the main text.

\section{Finite size effects in DMRG simulation}

Finding the interaction  between AFM bipolarons ($2H1S$ bound states) requires to solve the six-body $4H2S$ problem, 
which restricts the ED calculations to small system sizes .
To reach large enough system sizes, we use the DMRG algorithm ~\cite{white1992density,white1993density} on a
cylindrical lattice with open boundaries along the $x$ direction and periodic boundaries along the $y$ direction~\cite{gong2014emergent}. 
The lattice size is  $L_x$  along the $x$-direction and $L$ along the $y$-direction.

For the DMRG method to be applicable to our problem, the maximum values of $L_x$ and $L$ must be bigger than the linear size $\xi$ of the three-body bound state. 
The following verifications indicate that the DMRG method is indeed applicable to our problem.

\textit{(I) Comparing the ground state energy with ED.}

Tab.~\ref{tab:eg} provides a comparison between the DMRG and ED results on the same lattice size.
The difference between the energy values obtained from both approaches is negligibly small. 

\begin{table}[h!]
  \centering
  \caption{Comparison of ground state energy of  $2H1S$ system obtained from DMRG and ED.  The model parameters are: $t_1=1,t_3=-0.2,U=\infty$.}
  \label{tab:eg}
  \begin{tabular}{cccc}
    \toprule [0.08em]
    Lattice Size & $E_G^{ED} (t_1)$  & $E_G^{DMRG} (t_1)$ & Truncation error \\
    \midrule [0.08em]
    $8\times30$     & $-6.788045581$ & $-6.78452924$ & $0.33306691\times 10^{-15}$ \\
     $10\times30$  & $-6.757303233$ & $-6.75351219$ & $0.15909496\times 10^{-12}$ \\
    \bottomrule [0.08em]
  \end{tabular}
\end{table}

\textit{(II) Finite size scaling along the $L$ direction.}

\begin{figure}[!b]
\includegraphics[scale=0.23]{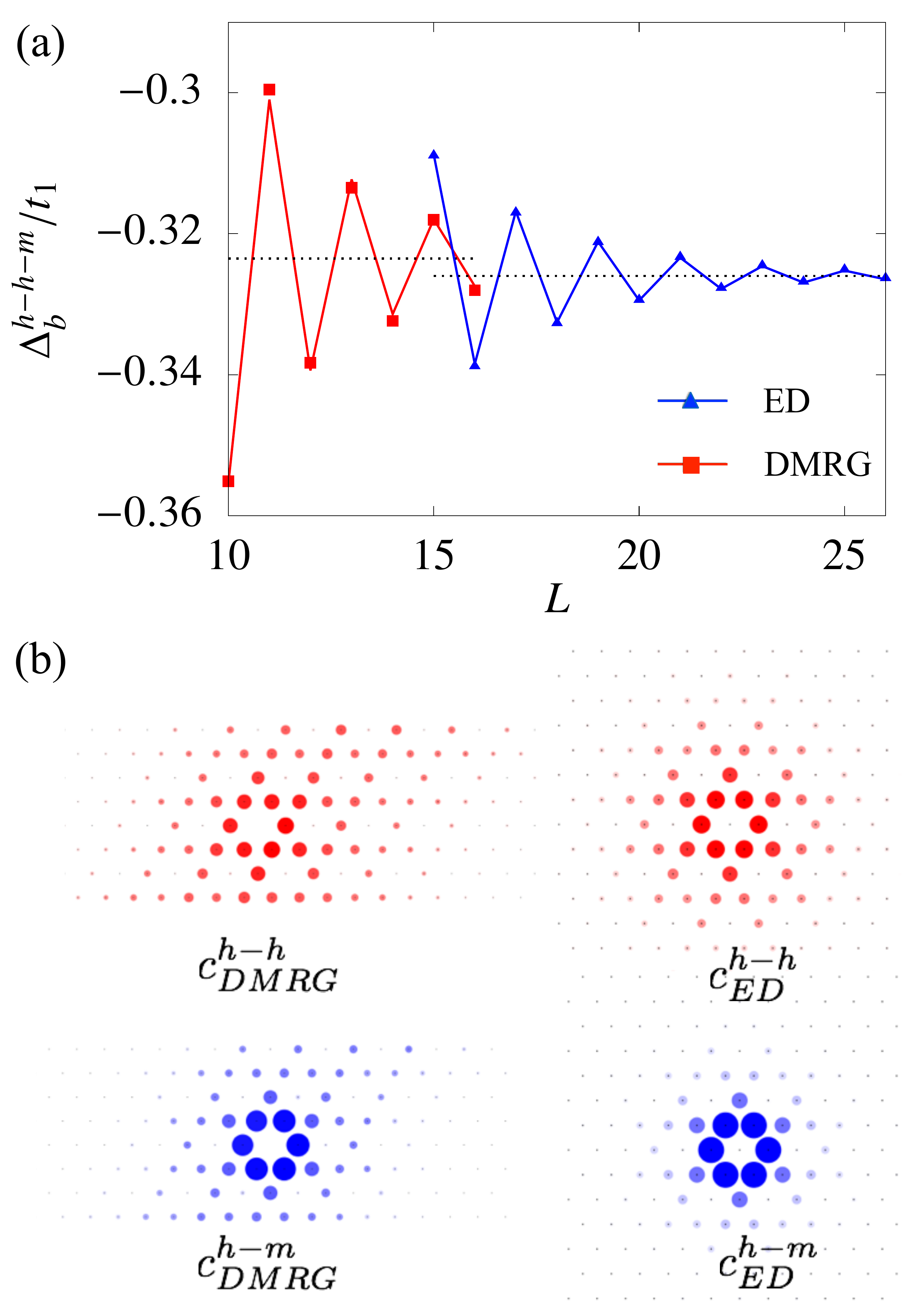}
\caption{(Color online) (a) Ground state energy obtained from DMRG (squares)  as a function of the strip width ($60\times L$) for $t_1=1,t_3=-0.2, U=\infty$ compared to
ED result (triangles) as a function of lattice size $L\times L$. The discrete points are numerical results for different sizes $L$, 
while the lines connecting the data points are  least-square fits to Eq.~\eqref{fit}. The horizontal lines are the extrapolations of $\Delta_{b}^{h-h-m}/t_1$ to the thermodynamic limit
$L \to \infty$ based on the fit of the DMRG and the ED results to Eq.~\eqref{fit}.
(b) Density-density correlation functions obtained from both DMRG and ED for the same model parameters as in (a) and lattice size $L_x\times L= 60\times 8$. The size and the color 
of the symbols are proportional to the values of the correlation functions.}
\label{fig:dmrg_sm}
\end{figure}

When using the DMRG method, the length $L_x$  can be made much larger than the size $\xi$ of the bound state. 
The main limitation arises from the width $L$. To verify that the finite width $L$ is not introducing a significant size effect, we performed a finite size scaling study of the ground state energy $E_G$ 
as a function of $L$. The results are shown in the Fig.~\ref{fig:dmrg_sm} (a). After fitting of the DMRG results with Eq.(\ref{fit}), we obtain $E_G/t_1=-0.32174\pm 0.00243$ within $95\%$ confidence interval, 
which is close to ED result: $-0.32591 \pm 1.1647\times 10^{-4}$ (Fig.~\ref{fig:dmrg_sm}(a)). 
Furthermore, Fig.~\ref{fig:dmrg_sm}(b) shows a comparison of various correlation functions obtained with DMRG and with ED. 
The correlation function matches with the ED result very well in spite of the limited size of the strip width, implying that the $2H1S$ bound state is practically not affected for the largest values of  $L$ that can be reached with DMRG. The asymmetry of the correlation function about the vertical line across the cloud center 
is due to the asymmetric open boundary condition of the finite triangular lattice  spanned by the primitive vectors: ${\bm a}_1=\hat{x}$ and 
${\bm a}_2=(\hat{x}-\sqrt{3}\hat{y})/2$.

\begin{figure}[!t]
\includegraphics[scale=0.6]{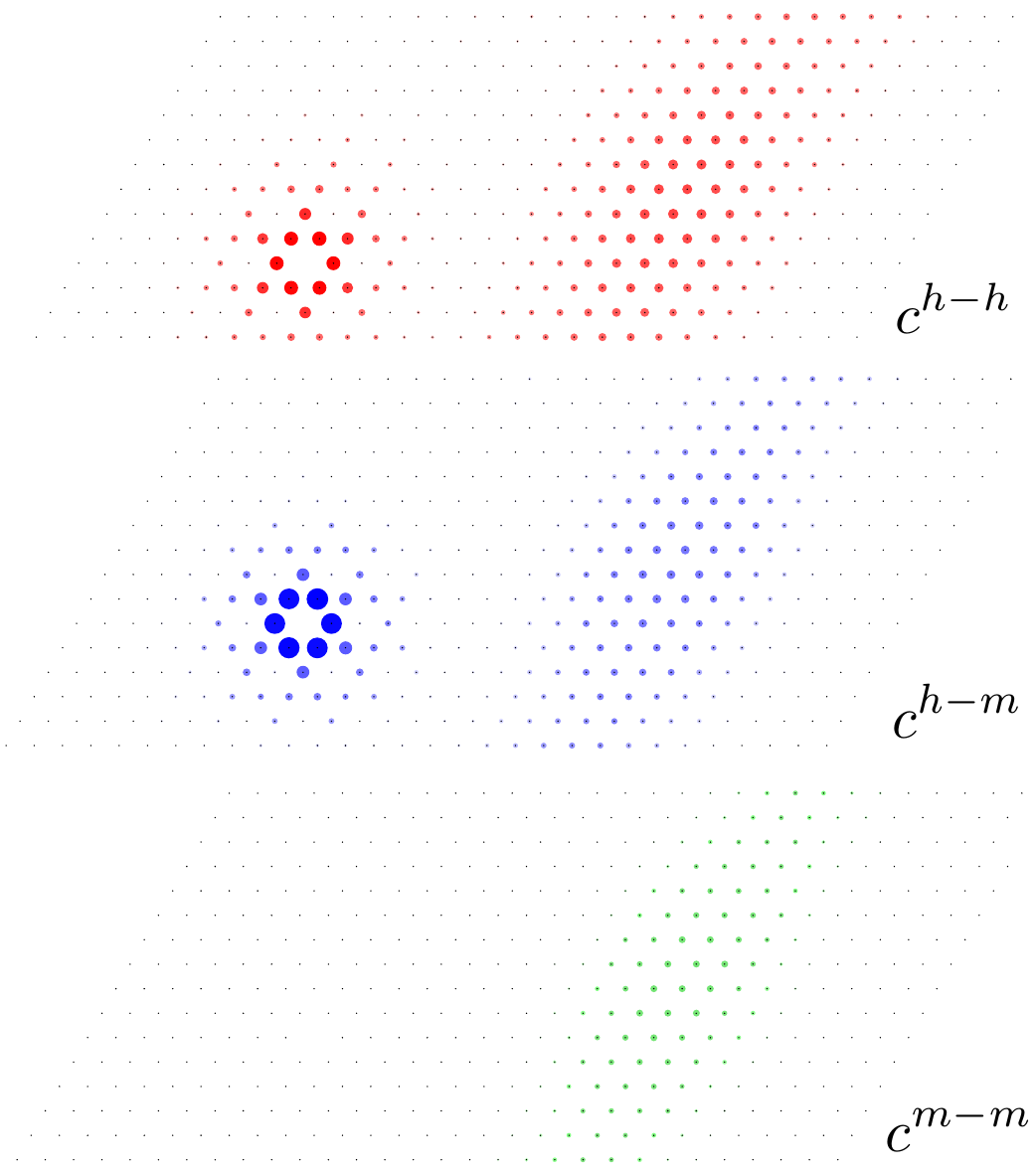}
\caption{(Color online) The different density-density correlation functions between holes and spin flips on a $16\text{(open)}\times 30\text{(periodic)}$ lattice for 4H2S system, 
which is obtained from DMRG simulation with truncation error about $10^{-7}$. The model parameters used here are $t_1=1,t_3=-0.2$ and $U=\infty$.}
\label{fig:dmrg_corr}
\end{figure}

\section{Density profile for four holes and two magnons ($4H2S$)}

\begin{figure}[!h]
\includegraphics[scale=0.25]{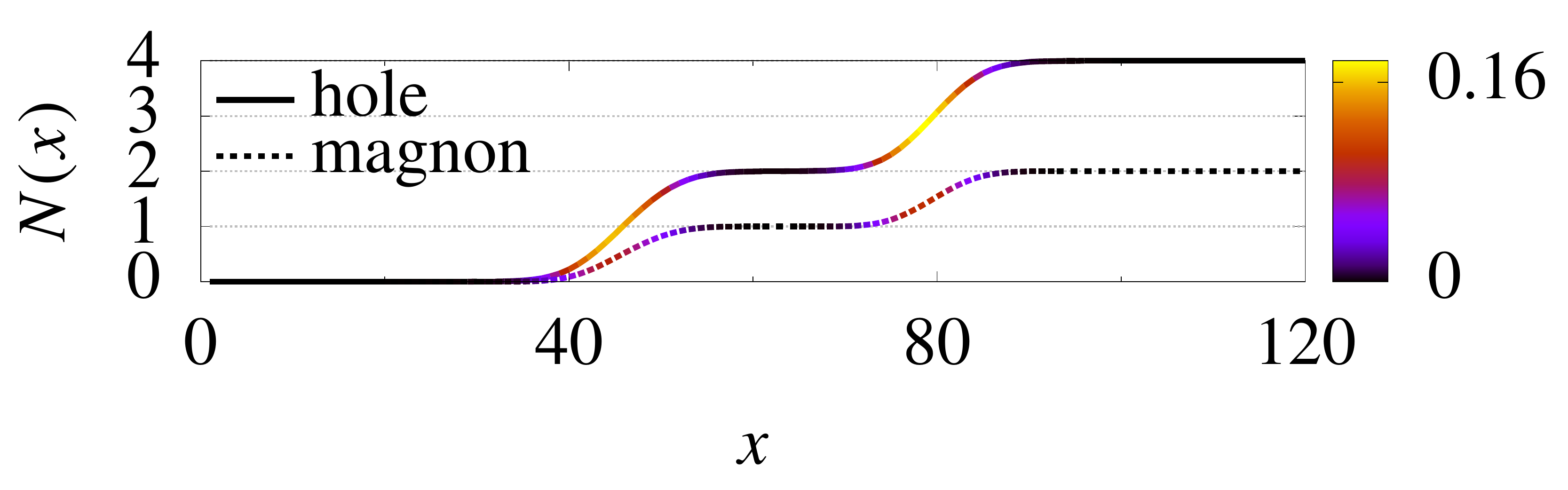}
\caption{(Color online) Accumulated particle number, $N(x)= \sum_{i<x} \sum_{j=1}^L \rho_{s,m}(i,j)$, for holes and magnons. The model parameters
are the same as in Fig.5(a) of the main text. The colored part of the curve (highest slope) indicates the location of the holes/magnon.}
\label{fig:dmrg_density}
\end{figure}

For the few-body problem, the DMRG algorithm works very well on a very wide stripe geometry, which is crucial for illustrating the formation and 
interaction of the bound states on a two-dimensional lattice. In this section, we extend the simulation of the 4H2S system to  a lattice of size $L_x\times L= 30 \times 16$.
The new results are consistent with those obtained from simulations of the $8\times 120$ strip shown by Fig.5(a).
Fig.~\ref{fig:dmrg_corr} shows the density-density correlation functions. 
Each 2H1S bound state has the same correlation functions obtained from ED calculation, which confirms the reliability of the DMRG simulation and 
further confirms the repulsion between AFM bipolarons.
The repulsive interaction between AFM bipolarons becomes even more transparent 
upon plotting the accumulated particle density along the x-direction, as shown in Fig. ~\ref{fig:dmrg_density}.
In agreement  with the correlation functions shown by Fig.~5(a) of the main text, 
the ground state consists of two well separated   thee-body bound states (AFM bipolarons) including two holes and one magnon.

\end{document}